\documentclass[]{elsarticle}
\usepackage{lineno,hyperref}
\usepackage{epstopdf} 
\usepackage{color}
\usepackage{lscape}
\modulolinenumbers[5]
\biboptions{sort&compress}
\usepackage{graphicx}
\usepackage{amssymb}
\begin{document}
\begin{frontmatter}
\title{Fast relaxation in metallic glasses studied by measurements of the internal friction at high frequencies}

\author[VSPU]{G.V. Afonin}
\author[NWPU]{J.C. Qiao}
\author[VSPU]{A.S. Makarov}
\author[IFTT]{N.P. Kobelev}
\author[VSPU]{V.A. Khonik}
\cortext[cor]{Corresponding author} 
\ead{v.a.khonik@yandex.ru} 
\address[VSPU] {Department of General Physics, Voronezh State Pedagogical
University,  Lenin St. 86, Voronezh 394043, Russia}
\address[NWPU] {School of Mechanics, Civil Engineering and Architecture, Northwestern Polytechnical University, Xi’an 710072, China}
\address[IFTT] {Institute of Solid State Physics RAS, Chernogolovka 142432, Russia}

\begin{abstract}
We performed high-frequency (0.4 to 1.7 MHz) measurements of the internal friction (IF) on 14 bulk metallic glasses (MGs). It is found that 12 of these MGs display relaxation IF peaks at temperatures $\approx 400$--500 K, which are weakly affected by heat treatment within the amorphous state. The corresponding relaxation time is about 0.3 $\mu$s. This fast relaxation is reported for the first time in the literature. 

The  apparent activation enthalpy for 4 MGs is determined. It is shown that if measured at a low frequency of 1 Hz, these IF peaks would appear near room temperature or by 30--50 K below it. Thus, the IF peaks found in the present investigation are direct analogues of low-temperature/low-frequency $\beta '$- and/or $\gamma$-relaxations described in the literature. 

It is argued that the origin of these IF peaks is related to the relaxation of structural defects similar to dumbbell interstitials, which are frozen in solid glass from the liquid state upon melt quenching. These defects can be considered as elastic dipoles and their re-orientation in the applied stress provides anelastic deformation and corresponding IF peak.  After correction for temperature dependence of the shear modulus, the true activation enthalpy (0.62 eV to 1.08 eV) and activation entropy (4$\leq \Delta S/k_B\leq 15$) for  defect re-orientation are determined. 

\end{abstract}

\begin{keyword}
metallic glasses, internal friction, relaxation, defects 
\end{keyword}

\end{frontmatter}

\section{Introduction}

It is long known that measurements of the anelasicity and/or viscoelasticity  constitute a very power tool to study internal processes occurring in solids on different spatial scales \cite{Nowick1972}. In particular, this type of testing provides important information upon applying sign-alternating stress, which allows to derive the measures of the dissipation of the mechanical energy  given by the internal friction (IF) (tangent of the phase lag) or the loss modulus. This way of testing is often applied to metallic glasses (MGs). Upon continued heating of MGs, a number of IF features are presently known. This is, first of all, so-called $\alpha$-relaxation, which is observed near the glass transition temperature $T_g$ and considered to be a manifestation of cooperative atomic movements leading to irreversible viscous flow \cite{HachenbergApplPhysLett2008,ZhaoJChemPhys2016,ShaoScrMater2023}. In the low-frequency domain ($10^{-2}$ Hz to 10 Hz) usually assessed by the dynamic mechanical spectroscopy (DMS) method, $\alpha$-relaxation is observed as a  separate damping peak near $T_g$ but at high  frequencies it is manifested as a rapid rise of the damping and subsequent sharp damping fall due to crystallization. At lower temperatures but still above room temperature,  MGs often exhibit a secondary $\beta$-relaxation observed  as a peak, shoulder or excess wing on DMS damping spectrum  also called as the Johari-Goldstein relaxation  \cite{YuMaterToday2013}. This $\beta$-relaxation is  linked to a number of still unclear issues in MGs related to defects, diffusion, fragility, mechanical, magnetic  and chemical properties \cite{YuMaterToday2013, ShaoScrMater2023,SunActaMater2016, DuanPRL2022,ZhouActaMater2023}. In addition, quite below room temperature (in the range 190 K $\leq T \leq$ 270 K  at frequencies of  several hundred Hz), a complicated large broad internal friction peak occurring due to MGs' plastic deformation was reported  \cite{KhonikActaMater1996}.   

Besides that, low frequency DMS testing of different MGs revealed another damping peak located also well below room temperature (in the range 150 K$\leq T \leq $250 K), which was named as the $\beta '$-relaxation \cite{WangNatCommun2015} or $\gamma$-relaxation \cite{KuchemannScrMater2017}. Sometimes it is said that $\beta '$ and $\gamma$ damping peaks correspond to separate relaxation processes \cite{ShaoScrMater2023}. While $\beta$-relaxation occurring above room temperature in the low frequency DMS domain is considered to be a slow process  \cite{ZhaoJChemPhys2016}, the $\beta '$- (and/or $\gamma$-) relaxation is thought to be related to certain fast processes \cite{ZhaoJChemPhys2016,WangNatCommun2015,WangMaterToday2017,ShaoScrMater2023,WangActaMater2020, WangMaterFutur2023}. The data on many MGs  show that $\beta '$-relaxation constitutes a universal process with a low activation energy of 0.3 eV to 0.6 eV, which can be related to the initiation of plasticity in MGs \cite{WangMaterToday2017}.  

It is generally accepted that MGs are non-uniform on a nanometer scale and can be considered as consisting of solid-like regions ("matrix") and isolated liquid-like regions ("defects") with low local viscosities and relaxation times  \cite{SunActaMater2016,WangMaterFutur2023,QiaoProgMaterSci2019}. By combining low-frequency DMS tests revealing low-temperature $\beta '$-relaxation and  molecular-dynamic simulations it was shown that the  fast dynamics of MGs at low temperatures is related to liquid-like regions, which are assumed to be inherited from the equilibrium liquid state. These liquid-like regions exhibit the behaviour similar to that in liquids, including comparable activation energies and low viscosities of $ \sim 10^7$ Pa$\times $s even below room temperature \cite{ChangNatureMater2022}.

In this work, we measured the internal friction (IF) in a number of  low-entropy- and high-entropy bulk MGs at high frequencies of 400 kHz to 1700 kHz upon heating up to the crystallization temperature. In most of the cases, we found an IF peak at a temperature of about 400 -- 500 K and present the evidence that the corresponding relaxation process is quite similar to the $\beta '$-relaxation revealed by low-frequency DMS tests near or below room temperature mentioned above. The origin of this relaxation is discussed.

\begin{table}[t]
\caption{\label{tab:table1} Metallic glasses under investigation, testing method (EMAT/RUS), internal friction peak temperatures $T_{max}$, which are observed at the frequencies $f_{max}$, and the glass transition temperatures $T_g$ measured at 3 K/min.  } 
\footnotesize
\begin{tabular}{p{3mm}|l|l|l|l|c}
\hline
\hline
No & Composition (at.\%)&Testing method&$T_{max}$(K)&$f_{max}$(kHz)&$T_g$ (K)\\ 
\hline
\hline

1 & Ti$_{20}$Zr$_{20}$Hf$_{20}$Cu$_{20}$Be$_{20}$ & EMAT& 405 & 595& 617\\
2 & Ti$_{20}$Zr$_{20}$Hf$_{20}$Ni$_{20}$Be$_{20}$& EMAT $\&$ RUS &392 & 530 & 619 \\
3 & Zr$_{31}$Ti$_{27}$Be$_{26}$Cu$_{10}$Ni$_{6}$ & RUS & 401 & 615& 597\\
4 & Ti$_{32.8}$Zr$_{30.2}$Be$_{22.7}$Cu$_{9}$Ni$_{6.3}$  & RUS & 373 & 546& 574\\
5 & Zr$_{35}$Cu$_{25}$Hf$_{13}$Al$_{11}$Ag$_{8}$Ni$_{8}$&EMAT& 433 & 578 & 710\\
6 & Zr$_{46}$Cu$_{46}$Al$_{8}$ &EMAT& 515 & 598& 688\\
7 & Zr$_{46}$Cu$_{45}$Al$_{7}$Ti$_{2}$ & EMAT $\&$ RUS & 470 & 456 & 682 \\
8 & Zr$_{48}$Cu$_{32}$Ag$_{8}$Al$_{8}$Pd$_{4}$ & EMAT & 508 & 580 & 672 \\
9 & Zr$_{50}$Cu$_{34}$Ag$_{8}$Al$_{8}$ & EMAT & 503 & 548 & 682 \\
10 & Zr$_{56}$Co$_{28}$Al$_{16}$ &EMAT& 467 & 641 & 731 \\ 
11 & Zr$_{57}$Nb$_{5}$Al$_{10}$Cu$_{15.4}$Ni$_{12.6}$ (Vit-106)& EMAT& 423 & 616& 665\\ 
12 & Zr$_{65}$Al$_{10}$Ni$_{10}$Cu$_{15}$ & EMAT& 379 & 576 & 627\\ 
13 & Pd$_{40}$Cu$_{30}$Ni$_{10}$P$_{20}$ & EMAT & -- &--& 560\\
14 & Pt$_{42.5}$Cu$_{27}$Ni$_{9.5}$P$_{21}$ & EMAT & -- &--& 511\\
\hline
\hline

\end{tabular}
\end{table}

\section{Experimental}

We studied 14 bulk Zr-, Ti-, Pd- and Pt-based MGs listed in Table 1. All MGs were
produced by melt suction method as 2 mm thick bars and confirmed by X-rays to be completely amorphous. All castings were also tested by differential scanning calorimetry (DSC), which showed quite usual thermal effects upon heating. The glass transition temperatures $T_g$ were determined as the onset of endothermal heat flow at 3 K/min. The glasses were further studied by two different methods, using electromagnetic acoustic transformation (EMAT) method and by the resonant ultrasound spectroscopy (RUS). EMAT method is based on the Lorentz interaction of the current in an exciting coil surrounding a sample with external permanent magnetic field  \cite{VasilievUFN1983}. By changing current frequency, one can perform frequency scanning and derive transverse  resonant vibration  frequency $f$ (400 kHz -- 600 kHz) of a sample ($5\times{5}\times{2}$ mm$^3$) by a pick-up coil. Frequency scanning was automatically performed every 10–15 s upon heating and the shear modulus was then calculated as $G(T)=G_{rt}f^2(T)/f^2_{rt}$, where $f_{rt}$ is the resonant frequency at room temperature and $G_{rt}$ is the room-temperature shear modulus taken from literature or determined by RUS. 

RUS testing was performed using a setup similar to that described in Ref.\cite{BalakirevRevSciInstrum2019} on $\approx 2\times 2\times 2$ mm$^3$ samples. The internal friction upon both EMAT and RUS measurements  was calculated as $Q^{-1}(T)=\Delta f(T)/f$, where $\Delta f$ is the width of the resonant curve at 0.707 of the resonant peak height and $f$ is the resonance frequency \cite{Nowick1972}. For both methods, the error of the absolute $G(T)$-values was accepted to be 1--2\% while the error in the measurements of the changes of this quantity was estimated to be $\approx 5$ ppm near room temperature and about 100 ppm near $T_g$. The background $Q^{-1}$ level (i.e. the IF uncertainty) in both methods was as low as 20-40 ppm at temperatures of about 400 K and by an order of magnitude larger near $T_g$. As matter of fact, this is unprecedented precision at such relatively high temperatures and frequencies. All measurements were performed in a vacuum of $\approx 0.01$ Pa at a heating rate of 3 K/min.

\begin{figure}[t]
\begin{center}
\includegraphics[scale=0.75]{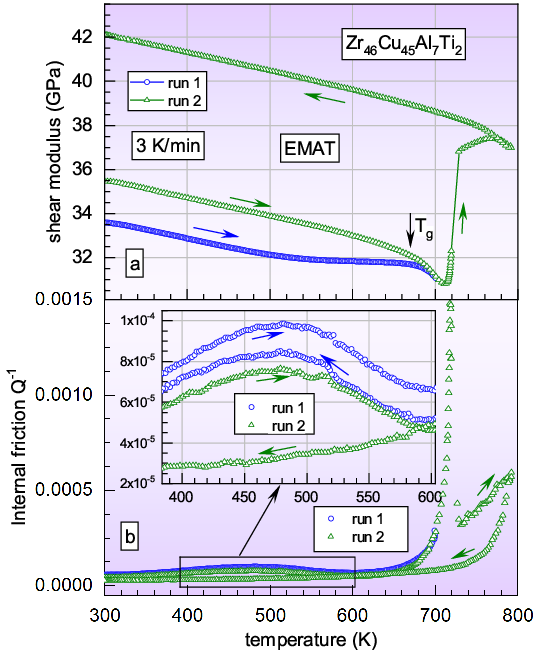}
\caption[*]{\label{Fig1.eps} Temperature dependences of the shear modulus (a) and internal friction (b)  of Zr$_{46}$Cu$_{45}$Al$_{7}$Ti$_{2}$ sample in the initial (run 1) and relaxed (run 2) states as determined by the EMAT method. The shear modulus $G_{rt}=33.6$ GPa is taken from Ref.\cite{WangJApplPhys2014}. Calorimetric glass transition $T_g$ temperature is indicated by the vertical arrow. Other arrows give the heating/cooling sequence. The inset in panel (b) shows the IF on an enlarged scale demonstrating an IF peak near 470 K, which disappears after crystallization.  }
\end{center}
\end{figure} 
  
\section{Results}
\subsection{Observation of high-frequency IF peaks}
Figure \ref{Fig1.eps} shows temperature dependences of the shear modulus $G$ (a) and internal friction $Q^{-1}$ (b) of glassy Zr$_{46}$Cu$_{45}$Al$_{7}$Ti$_{2}$. The sample was first heated into the supercooled liquid state (i.e. above $T_g\approx 670$ K) and cooled back to room temperature at the same rate (run 1). Next, it was heated up to 800 K that results in the complete crystallization and again cooled back to room temperature (run 2).  It is seen that temperature dependences of the shear modulus are quite usual (see e.g. Refs  \cite{MakarovIntermetallics2020,MitrofanovSciRep2016}) and include \textit{i}) an increase of $G$ (over its anharmonic decrease) due to exothermal structural relaxation  in the range 500 K$<T<650$ K seen in run 1, \textit{ii}) a decrease of $G$ above $T_g$, which is about the same during run 1 and run 2 and  \textit{iii}) a sharp $G$-rise due to  crystallization. The latter is accompanied by a very large IF peak ($Q^{-1}_{max}\approx 1.7\times 10^{-3}$) shown in panel (b) while the inset in this panel demonstrates a clear IF peak located at about 470 K with a  height of $Q^{-1}_{max}\approx 1\times 10^{-4}$. It is also seen that this height decreases by 20 -- 30 \% in the relaxed state (see the heating curve in run 2) but the peak completely disappears only after crystallization as demonstrated by the cooling curve in run 2.

\begin{figure}[t]
\begin{center}
\includegraphics[scale=0.75]{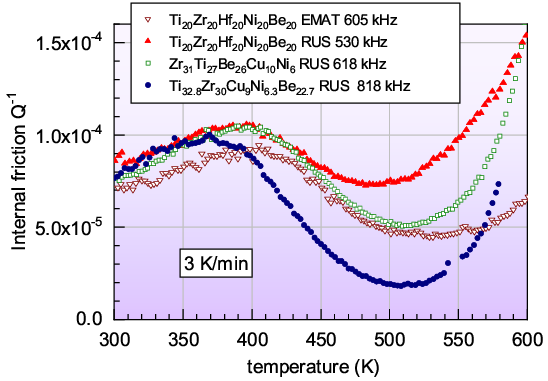}
\caption[*]{\label{Fig2.eps} Internal friction temperature dependences of indicated glasses tested by EMAT and/or RUS methods. An internal friction peak near 380 K to 400 K at indicated peak frequencies is seen. }
\end{center}
\end{figure}

Figure \ref{Fig2.eps} gives examples of this IF peak for three MGs. For high-entropy glassy Ti$_{20}$Zr$_{20}$Hf$_{20}$Be$_{20}$Ti$_{20}$, this Figure shows IF measurement results using both EMAT and RUS methods. It is seen that $Q^{-1}(T)$-dependencies are quite similar showing an IF peak at about $T_{max}\approx 400$ K, which has a height of $Q^{-1}_{max}\approx 1\times 10^{-4}$. The IF peaks for other glasses in Fig.\ref{Fig2.eps} have close $Q^{-1}_{max}$.

\begin{figure}[t]
\begin{center}
\includegraphics[scale=0.6]{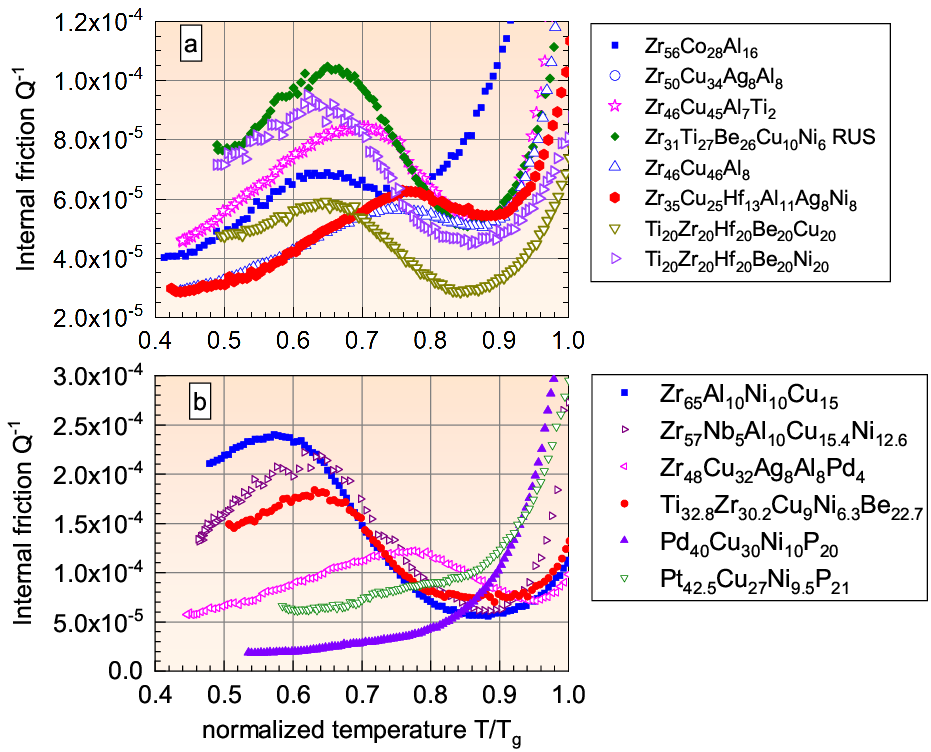}
\caption[*]{\label{Fig3.eps} Temperature dependences of the internal friction for all studied  metallic glasses tested at frequencies 400-600 kHz as a function of the normalized temperature $T/T_g$. Only a part of measured data points is shown. All glasses, except Pt- and Pd-based MGs, reveal internal friction peaks located in the range of $0.59\;T_g \leq T_{max} \leq 0.78 \;T_g$.}
\end{center}
\end{figure}

Figure \ref{Fig3.eps} gives temperature dependences of the  IF for all 14 glasses under study as a function of the normalized temperature $T/T_g$. It is seen that 12 MGs display an IF peak similar to those described above and located at temperatures $0.59\;\leq T_{max} \leq 0.78\;T_g$ ($T_g$-values are listed in Table 1). The height of these peaks $Q^{-1}_{max}$ changes from $\approx 5.7\times 10^{-5}$ (Fig.\ref{Fig3.eps}a) to $\approx 2.4\times 10^{-4}$ (Fig.\ref{Fig3.eps}b). It is to be noted that all peaks are well defined and there is no evidence of their merging with a sharp IF rise near $T_g$, which should be attributed to the beginning of $\alpha$-relaxation. It is also worthy of notice that the IF spectra differ quite a lot even for MGs, which are relatively close in chemical composition. In particular, $Q^{-1}(T)$-dependence of glassy Ti$_{20}$Zr$_{20}$Hf$_{20}$Be$_{20}$Ni$_{20}$ is pretty different from that for compositionally similar Ti$_{20}$Zr$_{20}$Hf$_{20}$Be$_{20}$Cu$_{20}$.

However, glassy Pt$_{42.5}$Cu$_{27}$Ni$_{9.5}$P$_{21}
$ and Pd$_{40}$Cu$_{30}$Ni$_{10}$P$_{20}$ do not display any clear IF peaks. The former one displays only small "remnants" of a peak in the range $0.7\;T_g\leq T \leq 0.8\; T_g$ while the latter glass shows only a smooth IF-rise with temperature without any signs of a peak despite of extremely high IF measurement resolution. Meanwhile, it is to be noted that low frequency DMS measurements do reveal an IF $\beta '$ relaxation peak in glassy Pd$_{40}$Cu$_{30}$Ni$_{10}$P$_{20}$ \cite{WangMaterToday2017}. The reason for this difference can be attributed to a specific IF frequency dependence in this particular glass.

Overall, one can conclude that high-frequency IF measurements show IF peaks at temperatures of about 400 K to 500 K at comparable frequencies of $\approx 500-600$ kHz. These IF peaks  are weakly affected  by relaxation performed by heating into the supercooled liquid state. At that, crystallization completely removes these peaks. To our knowledge, this is the first observation of high-frequency IF below $T_g$ peaks reported in the literature.

\subsection{Determination of the relaxation time}
While achievable changes of the resonant frequency in the EMAT method are very limited (usually 400 kHz to 600 kHz), RUS testing allows  IF measurements at different resonant modes and, therefore, provides a relatively wide frequency span. In the present investigation, detailed IF measurements were performed by RUS on 4 MGs listed in Table 2 in the range 500 kHz$\leq f\leq 1700$ kHz. Every glass was tested at 5-7 different frequencies in this range. It is found that the IF peak temperature $T_{max}$ increases with the frequency giving thus the evidence that this IF peak has relaxation nature. It is known that a relaxation IF peak occurs under the condition $2\pi f_{max}\tau=1$, where $f_{max}$ is the frequency related to an IF peak and  $\tau$ is the corresponding relaxation time \cite{Nowick1972}. Then, the IF peak frequencies $f_{max}$ listed in Table 1 correspond to the relaxation times $\tau\approx 0.3$ $\mu$s. The maximal IF peak frequency achieved in the present investigation is $1673$ kHz (see Fig.\ref{Fig4.eps}  below) that corresponds to a relaxation time of about 0.1$ \mu$s.  These relaxation times are very small indicating that the underlying relaxation process is very fast. 

For  a comparison, supercooled glassy Zr$_{65}$Al$_{10}$Ni$_{10}$Cu$_{15}$ (line 12 in Table 1) slightly above $T_g$ has a shear viscosity $\eta\approx 3\times 10^{10}$ Pa$\times$s and a shear modulus $G\approx 27$ GPa \cite{MakarovJPCM2021} and, therefore, the relaxation time $\tau= \eta/G$ is about 1 s. Another example can be taken from a detailed investigation \cite{ChangNatureMater2022}. For "liquid-like atoms" in glassy La$_{10}$Al$_{90}$ near room temperature the authors found a relaxation time of 0.7 ms. For such atoms at the same temperature in amorphous Y$_{68.9}$Co$_{31.1}$ they determined a shear modulus of 21.5 GPa and estimated a shear viscosity of $10^{6.44}$ Pa$\times$s (see supplemental information in this work) that corresponds to a relaxation time of $\approx 0.1$ ms. The relaxation times corresponding to the IF peak in the present work are smaller by three orders of magnitude. It should be emphasized that these relaxation times come directly from IF peak frequencies using the aforementioned condition for a relaxation IF peak and, therefore, are unquestionable.

 \bigskip
 
\section{Discussion}

\subsection{Apparent activation parameters and comparison with low frequency DMS data}

\begin{figure}[t]
\begin{center}
\includegraphics[scale=0.75]{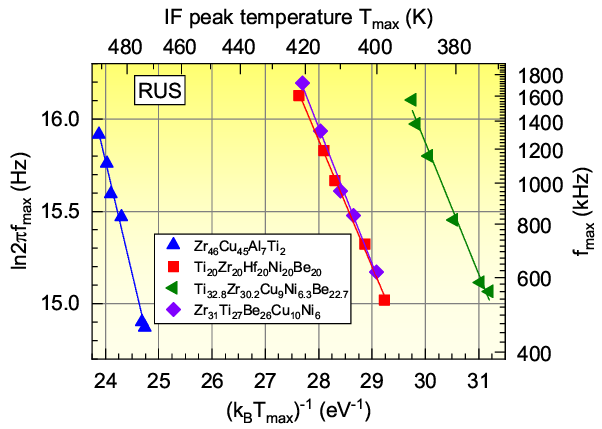}
\caption[*]{\label{Fig4.eps} Dependence of the frequency $f_{max}$ corresponding to the internal friction peak as a function of the peak temperature $T_{max}$ plotted according to Eq.(\ref{lnfmax}) for indicated MGs tested by RUS at different frequencies. The lines gives the least-square fits.  It is seen that this equation is obeyed very well. The errors along X- and Y-axes are close to the symbols' size. }
\end{center}
\end{figure}

Let us consider the relaxation process responsible for the IF peak in more details. The relaxation time $\tau$ corresponding to an IF peak at a given frequency $f_{max}$ is usually considered to follow the Arrhenius law, i.e. $\tau=\tau_0^{*}\;exp\left(\Delta H^*/k_BT\right)$, where $\Delta H^*$ is the apparent activation enthalpy, $\tau_0^{*}$ is a prefactor and $k_B$ is the Boltzmann's constant. This equation, together with the above condition for an IF peak, can be rewritten as

\begin{equation}
 ln (2\pi f_{max})=-ln\;\tau_0^{*}-\frac{\Delta H^*}{k_BT_{max}},
\label{lnfmax} \end{equation} 
where $T_{max}$ is the temperature of the IF peak observed at the frequency $f_{max}$. This relationship shows that $ln (2\pi f_{max})$ when plotted against the variable $(k_BT_{max})^{-1}$ should obey a straight line with the slope equal to the apparent activation enthalpy $\Delta H^*$. Figure \ref{Fig4.eps} shows this dependence for 4 MGs, each of which was tested by RUS at different frequencies. It is seen that Eq.(\ref{lnfmax}) for every glass is fulfilled very well proving its validity. The slopes of the linear fits in this Figure give the corresponding apparent activation enthalpies, which are given in Table 2.  It is seen that $\Delta H^*$ varies in the range from 0.69 eV to 1.23 eV for different MGs. 

It is interesting to note that the biggest $\Delta H^*$ is observed for glassy Zr$_{46}$Cu$_{45}$Al$_7$Ti$_2$, which  has the smallest entropy of mixing  $\Delta S_{mix}/R=-\sum\limits_{i=1}^nc_ilnc_i=0.98$ ($R$ is the universal gas constant, $c_i$ is the molar fraction of the \textit{i}-th element and $n$ is the number of constituent elements) while the smallest $\Delta H^*$ is found for Ti$_{20}$Zr$_{20}$Hf$_{20}$Be$_{20}$Ni$_{20}$, which has the biggest entropy of mixing $\Delta S_{mix}/R=1.61$ (see  Table 3 below). It was recently shown that the mixing entropy $\Delta S_{mix}$ is related to the excess entropy of \textit{solid} glass $\Delta S$ with respect to the maternal crystalline state \cite{AfoninAPL2024}, which can be determined calorimetrically. The so-called "high-entropy MGs"  (i.e. the glasses having high $\Delta S_{mix}$) actually constitute glasses with low excess entropy $\Delta S$ \cite{AfoninAPL2024}. While $\Delta S_{mix}$ reflects only the percentage of different atoms, the excess entropy $\Delta S$ accounts for all glass parameters (chemical composition, interatomic interaction, melt  quenching rate, entropy changes due to glass formation, etc.) \cite{AfoninAPL2024} and, besides that, describes the degree of glass disorder \cite{MakarovScrMater2024}. Therefore, glasses with the biggest $\Delta S_{mix}$ are most ordered, in line with the consideration by Yang et al. \cite{YangJPCL2022} given in terms of the "configurational entropy", which as a part includes $\Delta S_{mix}$.  Thus, the aforementioned dependence $\Delta H^*$($\Delta S_{mix}$) suggests that the most ordered MGs (i.e. having the biggest $\Delta S_{mix}$) should reveal the smallest $\Delta H^*$ and  vice versa. In turn, $\Delta S_{mix}$ can be related with polyamorphic transition in the supercooled liquid state \cite{YangActaMater2024}.  Further studies in this direction are desirable.

The apparent pre-exponential factor can be calculated as $\tau_0^*=exp(-A)$, where $A$ is the free term of the linear fits in Fig.\ref{Fig3.eps}. Table 2 shows that $\tau_0^*$-values for different MGs vary from $2\times 10^{-20}$ s to $6\times 10^{-16}$ s. At that, the smallest $\tau_0^*$ is obtained for the glass with the smallest mixing entropy $\Delta S_{mix}$ (line 1 in Table 2). 

\begin{table}[t]
\footnotesize
\caption{\label{tab:table2} Activation parameters of metallic glasses tested by RUS at various frequencies: apparent activation enthalpy $\Delta H^*$, linear fitting parameter   $A=-ln\;\tau_0^*$ (see Eq.(\ref{lnfmax})), IF peak temperature $T_{max}(f_1)$ for the frequency $f_1=1$ Hz  calculated using Eq.(\ref{Tmax}). } 

\footnotesize
\begin{tabular}{p{3mm}|l|l|l|l|l}
\hline
\hline
No & Composition (at.\%)&$\Delta H^*$ (eV)& $A=-ln\;\tau_0^*(s)$ & $\tau_{0}^*$ (s)&$T_{max}(f_1)$  (K)\\ 
\hline
\hline

1 & Zr$_{46}$Cu$_{45}$Al$_7$Ti$_2$ & 1.23$\pm 0.24$&$45.3\pm 5.6$ &$2\times 10^{-20}$ & $329\pm 20$ \\

2 & Zr$_{31}$Ti$_{27}$Be$_{26}$Cu$_{10}$Ni$_{6}$ & 0.74 $\pm 0.09$&$36.7\pm 2.7$&$1.2\times 10^{-16}$& $246\pm 8$ \\

3 & Ti$_{32.8}$Zr$_{30.2}$Cu$_{9}$Ni$_{6.3}$Be$_{22.7}$   & $0.71\pm 0.08$ & $37.4\pm 2.5$&$6\times 10^{-17}$& $233\pm 7$ \\ 

4 & Ti$_{20}$Zr$_{20}$Hf$_{20}$Be$_{20}$Ni$_{20}$ & 0.69$\pm 0.07$&$35.1\pm 2.2$ & $6\times 10^{-16}$ &$249\pm 7$  \\ 

\hline
\hline

\end{tabular}
\end{table}

To compare our IF data with the results of low-frequency DMS measurements discussed above, let us calculate the IF peak temperature $T_{max}(f_1)$ for a frequency  $f_1=1$ Hz, which is characteristic of DMS testing. For this purpose, Eq.(\ref{lnfmax}) can be rewritten as

\begin{equation}
\frac{1}{k_BT_{max}(f_1)}=\frac{1}{k_BT_{max}(f)}+\frac{ln\frac{f_1}{f}}{\Delta H^*}, \label{Tmax}
\end{equation}
where $T_{max}$ is the IF peak temperature, which is observed at a frequency $f$. 

The values of the IF peak temperatures $T_{max}(f_1)$ for the frequency $f_1=1$ Hz calculated with   $\Delta H^*$ determined for different MGs are given in Table 2. One can see  that if MGs displaying IF peaks at 400--500 K at frequencies of about 0.5 MHz (Table 1) are tested at a low frequency  $f_1=1$ Hz, they  should exhibit the same IF peaks near or quite below room temperature, at 230 K to 250 K. As mentioned above, such IF peak temperatures correspond to $\beta '$- and/or $\gamma$- relaxations. Low activation enthalpies $\Delta H^*$ (Table 2) are indicative of the same.          

\subsection{Possible relaxation mechanism and refinement of its activation parameters}
As shown earlier \cite{KobelevJALCOM2021}, the IF at temperatures below and near $T_g$ and frequencies ranging from $\approx 1$ Hz to $\approx 0.5$ MHz can be ascribed to the activation  of defects similar to dumbbell interstitials in crystalline metals, which are frozen-in from the melt upon glass production. These defects can be alternatively considered as elastic dipoles \cite{KobelevJApplPhys2014} and provide the  IF due to stress-induced \textit{i}) reversible transitions between equivalent energy states with a change of the dipole orientation and \textit{ii}) irreversible transitions from high-energy  into low-energy states. The latter mechanism provides irreversible IF changes upon temperature scanning. Meanwhile, the IF peaks reported in this work display  only a weak decrease of their height upon  heat treatment. This observation is in line with the estimates made for a number of MGs showing that only about one tenth of the whole defect number disappear due to structural relaxation upon annealing below or near $T_g$ \cite{KobelevJALCOM2021}. On the other hand, this finding indicates that the underlying relaxation processes are mostly reversible and can be ascribed to the aforementioned mechanism \textit{i}). Since the relaxation of interstitial-type defects (elastic dipoles) can quantitatively  explain quite a few relaxation phenomena in MGs from a common viewpoint (see Ref.\cite{KobelevUFN2023} for a detailed review), we believe that the understanding sketched above is most likely. As repeatedly mentioned in the literature, the application of the periodic external shear stress brings into motion about two dozens of atoms near the defect nucleus (see e.g. Ref.\cite{GranatoJNonCrystSol2006}). This provides  strong local shear softening \cite{GranatoJPCS1994} and possibly leads to the formation of the aforementioned "liquid-like regions" \cite{SunActaMater2016,ChangNatureMater2022,QiaoProgMaterSci2019}. The local shear softening, in turn, will contribute to the rise of the local internal friction. Such a phenomenon was recently documented in a detailed investigation \cite{YuNextMater2024}.

The assumption that the IF peaks under consideration are related to reversible  re-orientation of defects considered as elastic dipoles allows performing re-consideration of the activation parameters of the underlying relaxation process. This possibility first of all is related to the fact that the Arrhenius equation for the relaxation time has actually the form

 \begin{equation}
 \tau = \nu ^{-1}\;exp\left(\frac{\Delta H-T\Delta S}{k_BT}  \right), \label{tau} 
 \end{equation}
 where $\nu$ is the characteristic frequency of atomic vibrations ($\approx 10^{12}-10^{13}$ s$^{-1}$), $\Delta H$ is the true activation enthalpy and $\Delta S$ is the true activation entropy. Thus, the pre-exponential time constant is now determined as
 
 \begin{equation}
 \tau_0=\nu ^{-1}\;exp\left(-\frac{\Delta S}{k_B}\right), \label{tau0}
 \end{equation}
 i.e. strongly depends on the activation entropy. 
 
Second, one should account for the effect of temperature dependence of the activation enthalpy. It was earlier shown that the formal use of the Arrhenius equation (\ref{tau}) gives apparent values of the activation enthalpy and activation entropy, which turn out to be notably overestimated \cite{KobelevJETP2018}. The main physical origin for this consists in temperature dependence of the defect activation enthalpy, which is controlled by temperature dependence of the shear modulus\footnote{This statement is valid both for point defects in crystals \cite{KobelevJETP2018} and interstitial-type defects (elastic dipoles) in MGs \cite{KobelevUFN2023}.}.  The  true activation enthalpy  $\Delta H$ near room temperature and above it can be determined from the equation \cite{KobelevJETP2018}

\begin{equation}
\Delta H=\frac{\Delta H^*}{ \left[1+\frac{\partial ln(G)}{T\partial (\frac{1}{T})}\right]}\label{DeltaHapp},
\end{equation}
where $\Delta H^*$ is the apparent activation enthalpy determined from the standard Arrhenius equation and $G$ is the shear modulus. Because $G$ decreases with temperature, Eq.(\ref{DeltaHapp}) shows that $\Delta H$ is  smaller than $\Delta H^*$.
 
Since temperature dependences of the shear modulus were measured for all MGs under investigation, we performed corresponding estimates of the true activation enthalpies $\Delta H$ near the IF peaks using Eq.(\ref{DeltaHapp}), which are given in Table 3. It is seen that  $\Delta H$-values   are by 9--14\% smaller than the apparent activation enthalpies $\Delta H^*$ listed in Table 2. 

Besides that, accounting for the  same temperature dependence of the shear modulus leads to an additional coefficient in the expression (\ref{tau0}) for pre-exponential time constant, which now should be rewritten as 
\begin{equation}
\tau_0=\nu ^{-1}\;exp\left(-\frac{\Delta S}{k_B}\right)exp\left(\frac{\Delta H}{k_B}\frac{\partial lnG}{\partial T}\right). \label{tau0true}
\end{equation}

This equation shows that the consideration of $G(T)$-dependence  leads to a re-evaluation of the activation entropy. The estimates showed that the  increase of the apparent  activation entropy for MGs under investigation in $k_B$-units is from $\approx 2.2$ for Zr$_{46}$Cu$_{45}$Al$_7$Ti$_2$ up to  $\approx 3.3$ for other glasses. These values are close to the experimental uncertainty in the determination of $ln \;\tau_0^*$ (Table 2). 

Using Eq.(\ref{tau0true}),  one can now estimate the true activation entropy $\Delta S$. For this, we need to know the characteristic vibration frequency $\nu$. Since we assume that the relaxation process under discussion  is related to the re-orientation of interstitial-type elastic dipoles, it is natural to accept their characteristic vibration frequencies, which are equal to $(2\div 3)\times 10^{12}$ Hz \cite{KonchakovJPhysConMatt2019,KretovaJETPLett2020,KonchakovJETPLett2022}. This leads to $-ln(\nu^{-1})\approx28.5$.

\begin{table}[t]
\footnotesize
\caption{\label{tab:table3} Activation parameters of defect reorientation corrected for temperature dependence of the shear modulus (true activation enthalpy $\Delta H$ calculated using Eq.(\ref{DeltaHapp}) and true activation entropy $\Delta S/k_B$) together with the mixing entropy $\Delta S_{mix}$.} 

\begin{tabular}{p{3mm}|l|l|l|l}
\hline
\hline
No & Composition (at.\%)&$\Delta H$ (eV) & $\Delta S/k_B$ &$\Delta S_{mix}/R$\\ 
\hline
\hline

1 & Zr$_{46}$Cu$_{45}$Al$_7$Ti$_2$ & $1.08\pm 0.24$&$14.8\pm 5.5$&0.98 \\

2 & Zr$_{31}$Ti$_{27}$Be$_{26}$Cu$_{10}$Ni$_{6}$ &$0.67\pm 0.09$ &$5.2\pm 2.8$ &1.47 \\

3 & Ti$_{32.8}$Zr$_{30.2}$Cu$_{9}$Ni$_{6.3}$Be$_{22.7}$ &$0.63\pm 0.08$&$6.3\pm 2.4$&1.45  \\ 

4 & Ti$_{20}$Zr$_{20}$Hf$_{20}$Be$_{20}$Ni$_{20}$ & $0.62\pm 0.07$ &$3.6\pm 2.1$ &1.61  \\ 

\hline
\hline
\end{tabular}
\end{table}

Then, one obtains the true activation entropies $\Delta S$, which are listed in Table 3. For the MGs under investigation, the values of $\Delta S/k_B$  change from $\approx 4$ to $\approx 15$. Despite of relatively low accuracy of this calculation, this is, to our knowledge, the first experimental estimate of the activation entropy of defect re-orientation in MGs. It is  to be noticed in this connection that an independent estimate of the entropy per defect $\Delta S_{def}$ in MGs gives quite large values of 18 to 32 (in $k_B$-units), which are actually characteristic of dumbbell interstitial-type defects  \cite{MakarovJPCM2021b}. Although the activation entropy $\Delta S$ in the present investigation is not the same as and the entropy per defect $\Delta S_{def}$ considered in Ref.\cite{MakarovJPCM2021b}, the large and relatively close values of these quantities have to be mentioned. It is also worth noting that the smallest activation enthalpy and activation entropy are observed for the high-entropy glass with the largest entropy of mixing $\Delta S_{mix}$. This can evidence that glasses with bigger $\Delta S_{mix}$ are characterized by  more homogeneous energy relief. 

Finally, one should emphasize that experimental investigations of IF peaks in MGs in a wider frequency range should allow calculation of the activation parameters of defect re-orientation with higher accuracy. This will provide new valuable information on defects in MGs and  can also constitute a convenient instrument for the study of their internal energy relief.

\section{Conclusions}

We performed measurements of the shear modulus and internal friction (IF) at high frequencies in the range $0.4<f<1.7$ MHz on 14 bulk metallic glasses by using resonant ultrasound spectroscopy (RUS) and electromagnetic acoustic transformation (EMAT) testing methods from room temperature up to the temperatures of the complete crystallization. It is found that 12 MGs show relaxation IF peaks above room temperature but well below glass transition temperature and these peaks are not much affected by heat treatment within the glassy state. The relaxation times corresponding to the IF peaks are about 0.3 $\mu$s and, therefore, the underlying relaxation process is very fast. 

By performing RUS measurements at different frequencies on 4 MGs and assuming the Arrhenius relaxation law, the apparent activation enthalpies and pre-exponential factors for the IF peaks are determined. With this information, we showed that upon low-frequency DMS measurements ($\approx 1$ Hz) these peaks should appear near room temperature or by 30--50 K below it. Thus, the IF peaks found in the present investigation are direct analogues of  $\beta '$- and/or $\gamma$-relaxations reported in the literature for the case of low-frequency DMS testing.

It is argued that nature of the observed IF peaks is related  to stress-induced reversible relaxation of defects similar to dumbbell interstitials in crystalline metals, which  are frozen-in from the liquid state upon glass production. These defects can be considered as elastic dipoles and their re-orientation in sign-alternating external stress field leads to the anelastic relaxation and corresponding IF peak.

Assuming this relaxation mechanism, a re-consideration of the activation parameters is performed. The physical necessity  for this lies in the fact that the defect activation enthalpy depends on the shear modulus while the latter decreases with temperature. The experimental data on this temperature dependence derived in this work were used for a  re-estimation of the activation enthalpy and  pre-exponential factor. After all corrections we found that the true activation enthalpy  changes from 0.62 eV to 1.08 eV for the IF peaks in 4  MGs, which were subjected to a detailed study. The corresponding activation entropy varies from $\approx 15$ to $\approx 4$ (in $k_B$-units). To our knowledge, this is the first experimental estimate of the activation entropy for defect re-orientation in MGs.

\section{Acknowledgments}

The work was supported by the Russian Science Foundation under the project 23-12-00162.


\begin{thebibliography}{999}

\bibitem{Nowick1972} A.S. Nowick, B.S. Berry, Anelastic relaxation in crystalline solids, Academic Press, New York, London, 1972.

\bibitem{HachenbergApplPhysLett2008}  J. Hachenberg, D. Bedorf, K. Samwer, R. Richert, A. Kahl, M.D. Demetriou, W.L. Johnson, Merging of the $\alpha$ and $\beta$  relaxations and aging via the Johari-Goldstein modes in rapidly quenched metallic glasses, Appl. Phys. Lett. 92 (2008) 131911.

\bibitem{ZhaoJChemPhys2016} L.Z. Zhao, R.J. Xue, Z.G. Zhu, K.L. Ngai, W.H. Wang, H.Y. Bai, A fast dynamic mode in rare earth based glasses, J. Chem. Phys. 144 (2016) 204507.  

\bibitem {ShaoScrMater2023} L. Shao, L. Xue, J. Qiao, Q. Wang, Q. Wang, B. Shen, Gamma relaxation in Dy-based metallic glasses and its correlation with plasticity, Scr. Mater. 222 (2023) 115017.

\bibitem{YuMaterToday2013} H.-B. Yu, W.-H. Wang, K. Samwer, The $\beta$ relaxation in metallic glasses: an overview, Mater. Today 16 (2013) 183-191.

\bibitem{SunActaMater2016} B.A. Sun, Y.C. Hu, D.P. Wang, Z.G. Zhu, P. Wen, W.H. Wang, C.T. Liu, Y. Yang, Correlation between local elastic heterogeneities and overall elastic properties in metallic glasses, Acta Mater. 121 (2016) 266-276.

\bibitem{DuanPRL2022} Y.J. Duan, L.T. Zhang, J.C. Qiao, Y.-J. Wang, Y. Yang, T. Wada, H. Kato, J.M. Pelletier, E. Pineda, D. Crespo, Intrinsic correlation between the fraction of liquidlike zones and the $\beta$ relaxation in high-entropy metallic glasses, Phys. Rev. Lett. 129, 175501 (2022). 

\bibitem{ZhouActaMater2023} Z.-Y. Zhou, Y. Sun, L. Gao, Y.-J. Wang, H.-B. Yu, Fundamental links between shear transformation, $\beta$ relaxation, and string-like motion in metallic glasses, Acta Mater. 246 (2023) 118701.

\bibitem{KhonikActaMater1996} V.A.Khonik, L.V. Spivak, On the nature of low temperature internal friction peaks in metallic glasses, Acta Mater. 44 (1996) 367-381. 

\bibitem{WangNatCommun2015} Q. Wang, S.T. Zhang, Y. Yang, Y.D. Dong, C.T. Liu, J. Lu, Unusual fast secondary relaxation in metallic glass, Nature Commun. 6 (2015) 7876.

\bibitem{KuchemannScrMater2017} S. Küchemann, R. Maaß, Gamma relaxation in bulk metallic glasses, Scr. Mater. 137 (2017) 5–8. 

\bibitem{WangMaterToday2017} Q. Wang, J.J. Liu, Y.F. Ye, T.T. Liu, S. Wang, C.T. Liu, J. Lu, Y. Yang, Universal secondary relaxation and unusual brittle-to-ductile transition in metallic glasses, Mater. Today 20 (2017) 293-300.

\bibitem{WangActaMater2020} B. Wang, L.J. Wang, B.S. Shang, X.Q. Gao, Y. Yang, H.Y. Bai, M.X. Pan, W.H. Wang, P.F. Guan, Revealing the ultra-low-temperature relaxation peak in a model metallic glass, Acta Mater. 195 (2020) 611–620.   


\bibitem{WangMaterFutur2023} Q. Wang, Yi.-H. Shang, Y. Yang, Quenched-in liquid in glass, Mater. Futures 2 (2023) 017501.

\bibitem{QiaoProgMaterSci2019} J.C. Qiao, Q. Wang, J.M. Pelletier, H. Kato, R. Casalini, D. Crespo, E. Pineda, Y. Yao, Y. Yang, Structural heterogeneities and mechanical behavior of amorphous alloys, Prog. Mater. Sci. 104 (2019) 250–329.

\bibitem{ChangNatureMater2022} C. Chang, H.P. Zhang, R. Zhao, F.C. Li, P. Luo, M.Z. Li, H.Y. Bai, Liquid-like atoms in dense-packed solid glasses, Nature Mater. 21 (2022) 1240-1245.

\bibitem{VasilievUFN1983} A.N. Vasil'ev, Yu.P. Gaidukov, Electromagnetic excitation of sound in metals, Sov. Phys. Usp. 26 (1983) 952-973.

\bibitem{BalakirevRevSciInstrum2019} F.F. Balakirev, S.M. Ennaceur, R. J. Migliori, B. Maiorov, A. Migliori, Resonant ultrasound spectroscopy: the essential toolbox, Rev. Sci. Instrum. 90 (2019) 121401.

\bibitem{WangJApplPhys2014} D.P. Wang, D.Q. Zhao, D.W. Ding, H.Y. Bai, W.H. Wang,  Understanding the correlations between Poisson’s ratio and plasticity based on microscopic flow units in metallic glasses, J. Appl. Phys. 115 (2014) 123507.

\bibitem{MitrofanovSciRep2016} Y.P. Mitrofanov, D.P. Wang, A.S. Makarov, W.H. Wang, V.A. Khonik, Towards understanding of heat effects in metallic glasses on the basis of macroscopic shear elasticity, Sci. Rep. 6 (2016) 23026.

\bibitem{MakarovIntermetallics2020} A.S Makarov, Yu.P. Mitrofanov, E.V. Goncharova, J.C. Qiao, N.P. Kobelev, A.M. Glezer, V.A. Khonik, Relationship between the shear moduli of metallic glasses and their crystalline counterparts, Intermetallics 125 (2020) 106910.  

\bibitem{WangMaterToday2017} Q. Wang, J.J. Liu, Y.F. Ye, T.T. Liu, S. Wang, C.T. Liu, J. Lu, Y. Yang, Universal secondary relaxation and unusual brittle-to-ductile transition in metallic glasses, Mater. Today 20 (2017) 293-300.

\bibitem{MakarovJPCM2021} A.S. Makarov, J.C. Qiao, N.P. Kobelev,  A.S. Aronin,  V.A. Khonik, Relation of the fragility and heat capacity jump in the supercooled liquid region with the shear modulus relaxation in metallic glasses, J. Phys. Cond. Matter 33 (2021) 275701. 

\bibitem{AfoninAPL2024} G.V. Afonin, J.C. Qiao, A.S. Makarov, R.A. Konchakov, E.V. Goncharova, N.P. Kobelev, V.A. Khonik, High entropy metallic glasses, what does it mean? Appl. Phys. Lett. 124 (2024) 151905.

\bibitem{MakarovScrMater2024} A.S. Makarov, G.V. Afonin, R.A. Konchakov, V.A. Khonik, J.C. Qiao, A.N. Vasiliev, N.P. Kobelev, Dimensionless parameter of structural ordering and excess entropy of metallic and tellurite glasses, Scr. Mater. 239 (2024) 115783. 

\bibitem{YangJPCL2022} M. Yang, W. Li, X. Liu, H. Wang, Y. Wu, X. Wang, F. Zhang, Q. Zeng, D. Ma, H. Ruan, Z. Lu, Configurational entropy effects on glass transition in metallic glasses, J. Phys. Chem. Lett. 13 (2022) 7889-7897.

\bibitem{YangActaMater2024} Q. Yang, X.-M. Yang, T. Zhang, X.-W. Liu, H.-B. Yu, Structure and entropy control of polyamorphous transition in high-entropy metallic glasses, Acta Mater. 266 (2024) 119701. 

\bibitem{KobelevJALCOM2021} N.P.Kobelev, J.C. Qiao, A.S. Makarov, A.M. Glezer, V.A. Khonik, Internal friction and dynamic shear modulus of a metallic glass in a seven-orders-of-magnitude frequency range. J. Alloys Comp. 869 (2021) 159275.

\bibitem{KobelevJApplPhys2014} N.P. Kobelev, V.A. Khonik, A.S. Makarov, G.V. Afonin, Yu.P. Mitrofanov, On the nature of heat effects and shear modulus softening in metallic glasses: a generalized approach, J. Appl. Phys. 115 (2014) 033513.

 \bibitem{KobelevUFN2023} N.P. Kobelev, V.A. Khonik, A novel view of the nature of formation of metallic glasses, their structural relaxation, and crystallization, Physics -- Uspekhi 66 (2023) 673-690.
 
\bibitem{GranatoJNonCrystSol2006} A.V. Granato, A comparison with empirical results of the interstitialcy theory of condensed matter, J. Non-Cryst. Sol. 352 (2006) 4821-4825. 

\bibitem{GranatoJPCS1994} A.V. Granato. Self-interstitials as basic structural units of liquids and glasses,  J. Phys. Chem. Sol. 55 (1994) 931-939.

\bibitem{YuNextMater2024} H.-B. Yu, Q. Wang, Liquid-like clusters in glassy solids as a unique state of matter: dissipative but non-diffusive, Next Mater. 3 (2024) 100168.
 
\bibitem{KobelevJETP2018} N.P. Kobelev, V.A. Khonik, On the enthalpy and entropy of point defect formation in crystals, J. Exp. Theor. Phys. 126 (2018) 340-346. 

\bibitem{KonchakovJPhysConMatt2019} R.A. Konchakov, A.S. Makarov, N.P. Kobelev, A.M. Glezer, G. Wilde, V.A. Khonik, Interstitial clustering in metallic systems as a source for the formation of the icosahedral matrix and defects in the glassy state, J. Phys.: Condens. Matter 31 (2019) 385703.

\bibitem{KretovaJETPLett2020} M.A. Kretova, R.A. Konchakov, N.P. Kobelev, V.A.Khonik,  Point defects and their properties in the Fe$_{20}$Ni$_{20}$Cr$_{20}$Co$_{20}$Cu$_{20}$ high-entropy alloy, J. Exp. Theor. Phys. Lett. 111 (2020) 679–684.

\bibitem{KonchakovJETPLett2022} R.A. Konchakov, A.S. Makarov, A.S. Aronin, N.P. Kobelev, V.A. Khonik, Elastic dipoles in crystal and glassy aluminum and high-entropy Fe$_{20}$Ni$_{20}$Cr$_{20}$Co$_{20}$Cu$_{20}$ alloy, J. Exp. Theor.Phys. Lett.  115 (2022) 280–285.    

\bibitem{MakarovJPCM2021b}  A.S. Makarov, G.V. Afonin, J.C. Qiao, A.M Glezer, N.P. Kobelev, V.A. Khonik, Determination of the thermodynamic potentials of metallic glasses and their relation to the defect structure, J. Phys.: Condens. Matter 33 (2021)  435701. 


\end{thebibliography}
\end{document}